\documentclass[twocolumn,epsfig,superscriptaddress]{revtex4-1}

\usepackage{dcolumn}
\usepackage{bm}
\usepackage{amssymb}
\usepackage{subfigure}
\usepackage{here}
\usepackage{ulem}
\usepackage{enumerate}
\usepackage{cancel}
\usepackage{flushend}
\usepackage{bbold}
\usepackage{mathtools}
\usepackage{float}
\usepackage{stmaryrd}
\usepackage{colortbl}
\usepackage[table]{xcolor}
\usepackage{array,ragged2e}
\usepackage{amsmath,amssymb,amsfonts,stmaryrd,wasysym,graphicx,multirow,color,textcomp,subfigure}
\usepackage{url}
\usepackage[colorlinks=true,citecolor=blue,urlcolor=blue]{hyperref}
\usepackage{romannum}
\usepackage{tabularx}
\usepackage{changes}

\usepackage[utf8]{inputenc}
\usepackage{tikz-cd}

\usepackage{graphicx}
\usepackage{color}

\newenvironment{helv}{\fontfamily{phv}\selectfont}{\par}
\DeclareTextFontCommand{\texthelv}{\helv}

\makeatletter
\newcommand{\thickhline}{%
    \noalign {\ifnum 0=`}\fi \hrule height 1pt
    \futurelet \reserved@a \@xhline
}
\newcolumntype{"}{@{\hskip\tabcolsep\vrule width 1pt\hskip\tabcolsep}}
\makeatother

\begin{document}

\draft
\title{Double layered transformation cavities: Crossover between local and global transformations}
\author{Jung-Wan Ryu}
\email{jungwanryu@gmail.com}
\address{Center for Theoretical Physics of Complex Systems, Institute for Basic Science (IBS), Daejeon 34126, Republic of Korea}
\date{\today}

\begin{abstract}
We study the resonant modes in double-layered transformation cavities consisting of inner and outer layer boundaries where the mode intensity is mainly located inside the inner layer. This is an intermediate design between a transformation cavity and a cavity in a wholly transformed space of transformation optics. We demonstrate the crossover between these two extreme cases as the outer layer of the cavity varies and also explore the properties of the resonant modes in the cavity. While the near-field patterns of the resonant modes do not change as the outer layer becomes larger, the Q-factors approach those of a cavity in the wholly transformed space of transformation optics, and the far-field patterns are modified.
\end{abstract}

\maketitle

\section{Introduction}

Transformation optics (TO) provides a general design method to manipulate electromagnetic waves in a desired manner \cite{Leo06,Pen06}. It is based on the concept that Maxwell’s equations can be form-invariant under coordinate transformations, such that only the permittivity and permeability tensors are modified \cite{Pos62,Jac75}. Consequently, TO has become a powerful tool to control the propagation of electromagnetic waves \cite{Che10-2, Kil11, Liu12}. Many novel optical devices and unprecedented applications have been demonstrated with TO, including individual cloaks \cite{Pen06,Sch06,Cai07,Lan13}, wave concentrators \cite{Jia08,Sad15}, optical black holes \cite{Gen09,Che10,Deh15}, waveguide devices \cite{Rob08,Zha09,Tic10}, and negative refractive index materials \cite{Ves68,She01,Zha16}.

Transformation optics can be applied not only to the control of the propagation path of electromagnetic waves in a desired manner but also to the manipulation of the resonant mode properties of two-dimensional dielectric optical cavities \cite{Kim16}. One new kind of microcavity designed utilizing conformal TO to manipulate the resonant mode properties is called a transformation cavity (TC), which has a deformed boundary shape and inhomogeneous profile of the refractive index inside the microcavity associated with a given conformal mapping. The variation in the Q-factors and emission directionality of resonant modes in the system parameter space was systematically studied to find the optimized resonant modes in a lima\c{c}on-shaped cavity \cite{Ryu19}. The applicability of TCs can be further extended to arbitrary shapes by invoking a quasi-conformal mapping method \cite{Par19}. It has been expected that metamaterials, e.g., subwavelength structures such as holes and posts that can produce homogenized effective media, is one of the promising candidates for the implementation of TCs, like in numerous applications of TO theory \cite{Ves68, Pen99, Smi00, Smi04, Eng06, Val09, Gab09, Vas10, Gao12}. 

While space is globally transformed by a given conformal mapping in the case of original TO, space is locally transformed by mapping only inside the cavity in the case of TCs. Accordingly, the refractive indices inside and outside cavities have inhomogeneous profiles in the case of original TO but have inhomogeneous and homogeneous profiles, respectively, in the case of TCs. This discrepancy results in significant modifications of the resonant modes, thereby enabling us to independently control optical properties such as the Q-factor and emission directionality of the conformal whispering gallery mode (cWGM) in a TC. In this work, we propose a double-layered transformation cavity (DLTC) consisting of inner and outer layers with gradually varying refractive index profiles, which is an intermediate design between a TC and a cavity in a wholly transformed (WT) space of TO. We study the optical properties, namely the Q-factors and near- and far-field patterns, of the resonant modes in the DLTC. In the cavity, as the outer layer boundary becomes larger while the inner layer boundary does not change, the Q-factors of the resonant modes approach those of a cavity in WT space, and the emission directionalities change only by modifying the outer layer boundary shape.

\section{Systems: double layered transformation cavity}
\label{sec:systems}
{\it Deformed cavity and resonant modes -}
We consider a two-dimensional dielectric cavity with a given boundary shape, where the optical modes are described by the resonant modes. The modes are obtained by solving the following scalar wave equation,
\begin{equation}
[\nabla^2 + n^2 (x,y) k^2] \psi (x,y) = 0,
\end{equation}
where $n(x,y)$ is the refractive index function at the position $(x,y)$. The resonant modes should satisfy the outgoing-wave boundary condition, $\psi(x,y) \sim h_k(\theta)e^{i k r} / \sqrt{r}$, where $(x,y)=(r \cos\theta, r \sin\theta)$ and $h_k(\theta)$ is the far-field angular distribution of the emission. In a deformed cavity with a homogeneous refractive index profile, $n(x,y)$ is $n_0$ inside and $1$ outside the cavity. The real and imaginary parts of the complex wave number $k$ are, respectively, $\omega / c$ where $\omega$ is the frequency of the resonant mode and $c$ is the speed of light, and $- 1/(2 c \tau)$ where $\tau$ is the lifetime of the mode. The quality factor $Q$ of the mode is defined as $Q = 2 \pi \tau / T = - \mathrm{Re}(k) / 2 \mathrm{Im}(k)$. Here, we focus on transverse magnetic (TM) polarization, of which the wave function $\psi(x,y)$ corresponds to $E_z$, the $z$ component of the electric field \cite{Jac75}. In the case of TM polarization, both the wave function $\psi(x,y)$ and its normal derivative $\partial_\nu \psi$ are continuous across the cavity boundary.

\begin{figure}[tb]
\centering
\includegraphics[width=\linewidth]{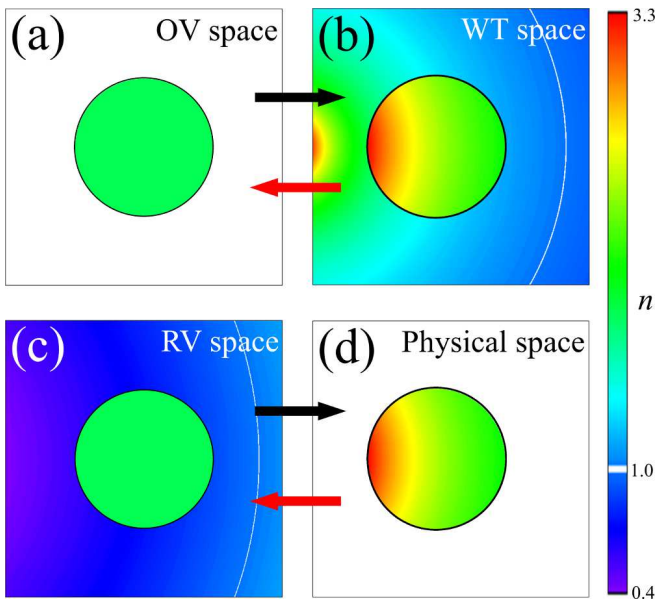}
\caption{Boundary shapes (black) and refractive indices (scaled color) for (a) a circular cavity in an original virtual space, (b) a lima\c{c}on-shaped cavity in a wholly transformed space, (c) a circular cavity in a reciprocal virtual space, and (d) a transformation cavity in a physical space. The black and red arrows denote the conformal mapping of Eq.~(\ref{conformal}) and its inverse, respectively. The white regions represent a refractive index of $n=1$.}
\label{fig0}
\end{figure}

{\it Four spaces and the difference between WT and physical spaces -} There are four spaces to understand the relations between original TO theory and a TC in depth: an original virtual (OV) space or a reference space  [Fig.~\ref{fig0} (a)], a WT space corresponding to the physical space of conventional TO theory [Fig.~\ref{fig0} (b)], a physical space for a TC [Fig.~\ref{fig0} (d)], and a reciprocal virtual (RV) space [Fig.~\ref{fig0} (c)]. We first consider a circular cavity in an OV space. Here, the cavity has a homogeneous refractive index, and the resonant modes are the solution of Maxwell's equations in dielectric media. Maxwell's equations remain invariant under a given conformal mapping, by which an OV space is transformed into a WT space that corresponds to the physical space of conventional TO theory. In a WT space, the cavity boundary is deformed and the inhomogeneous profile of a refractive index in a whole space can be determined by the ratio between the local length scales in both spaces. To make a TC in a physical space, the refractive index outside the cavity is set to $1$, that is, the conformal mapping is applied not only to the boundary deformations but also to the inside of the cavity, unlike in WT space where the mapping is applied to the whole space. These characteristic settings for TCs result in novel properties of the resonant modes in TCs. Besides, an RV space is given by an inverse conformal mapping from the physical space, in which refractive indices have homogeneous and inhomogeneous profiles inside and outside the cavity, respectively. Although RV space was originally introduced for numerical calculations such as boundary element methods \cite{Kim16,BEM} and Husimi functions \cite{Kim18}, which are used in this work, the space is useful to understand the resonant mode properties since the interior of a cavity in RV space is the same as that of a cavity in OV space. While the resonant modes obtained from OV and corresponding WT spaces are intrinsically the same, modes in WT and corresponding physical spaces are different in principle.

\begin{figure}[tb]
\centering
\includegraphics[width=\linewidth]{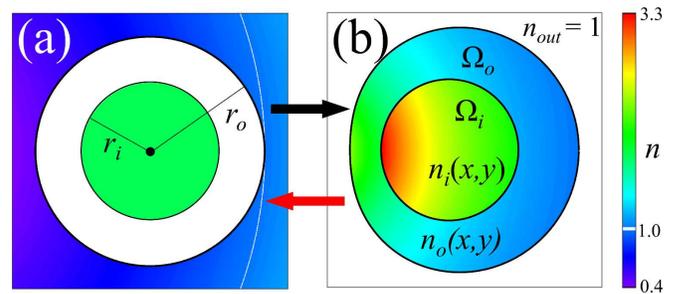}
\caption{Boundary shapes (black) and refractive indices (scaled color) for (a) a concentric circular layered cavity in a reciprocal virtual space and (b) a double-layered transformation cavity in a physical space. The regions of the inner and outer layers are $\Omega_i$ and $\Omega_o$, respectively, and the refractive indices of the inner and outer layers are $n_i (x,y)$ and $n_o (x,y)$, respectively.}
\label{fig1}
\end{figure}

{\it Double-layered lima\c{c}on-shaped transformation cavity -} In WT space, conformal mapping is applied to the whole space [Fig.~\ref{fig0} (b)], but in physical space, conformal mapping is applied only to the inside of the cavity [Fig.~\ref{fig0} (d)]. In order to bridge the gap between these two limiting cases, we propose a second layer outside of the inner layer. The resulting cavity has two sub-regions with inhomogeneous refractive index profiles: inner and outer layers, denoted by $\Omega_i$ and $\Omega_o$, respectively, as shown in Fig.~\ref{fig1} (b). As an example, we consider a DLTC with boundaries given by a lima\c{c}on shape, which is one of the widely studied shapes in the fields of quantum billiards \cite{Rob83}, deformed microcavities \cite{Wie08}, and transformation cavities \cite{Kim16}. The corresponding conformal mapping from the unit circle to the lima\c{c}on shape is given by
\begin{equation}
\zeta = \beta (\eta + \epsilon \eta^2),
\label{conformal}
\end{equation}
where $\eta = u + i v$ and $\zeta = x + i y$ are complex variables denoting positions in virtual spaces, i.e., OV and RV spaces, and physical spaces (including WT spaces), respectively. $\beta$ is a positive size-scaling parameter and $\epsilon$ is a deformation parameter. The inner and outer layer boundaries are given by
\begin{eqnarray}
x &=& \beta(r \cos(\phi) + \epsilon r^2 \cos(2 \phi)),\\\nonumber
y &=& \beta(r \sin(\phi) + \epsilon r^2 \sin(2 \phi)),
\label{boundary}
\end{eqnarray}
where size scale factors $r$'s are $r_i$ and $r_o$ for the inner and outer layer boundaries, which are equivalent to the radii of two circles of a concentric circular layered cavity in an RV space [Fig.~\ref{fig1} (a)], respectively, and $\beta = \beta_{max} = 1/(1+2\epsilon)$. The maximal value $\beta_{max}$ of $\beta$, that supports the cWGMs in a limaçon-shaped TC with $r=1.0$, can be obtained from the condition $|d{\zeta} / d{\eta} |^{-1}  \geq 1$ necessary for total internal reflection in the TC with outside refractive index $n_{out} = 1$. The refractive index profiles are given by
\begin{eqnarray}
n(x,y) =
\begin{cases}
 n_0 \left| \frac{d \zeta}{d \eta} \right|^{-1} = \frac{n_0}{\beta\left|\sqrt{1+4 \epsilon \zeta / \beta}\right|}, ~& (x, y) \in \Omega_i \\
 \left| \frac{d \zeta}{d \eta} \right|^{-1} = \frac{1}{\beta\left|\sqrt{1+4 \epsilon \zeta / \beta}\right|}, ~& (x, y) \in \Omega_o \\
 1, ~& \mathrm{for~air} 
\end{cases}
\label{indices}
\end{eqnarray}
where $n_0$ is the refractive index of the disk cavity in the OV space. In this DLTC, the boundary shapes of the inner and outer layers are lima\c{c}on, and the refractive index ratio at the interface between them is equivalent to $n_0$ irrespective of boundary positions. It is noted that we use the parameter $r_o \le 1.0$ for the outer layer boundary in order to avoid the local refractive index being smaller than $1.0$.

\section{Results}

{\it Resonant modes -}
We obtained the resonant modes in a DLTC in which both inner and outer layer boundary shapes are lima\c{c}on. The DLTC corresponds to a concentric circular layered cavity with $r=r_i$ and $r=r_o$ when $\epsilon=0$, where there are whispering gallery modes (WGMs) circling the inner layer boundary supported by the total internal reflection at the interface between the inner and outer layers. We start from the WGM with mode numbers $(m,l)=(13,1)$, where $m$ is the azimuthal mode number and $l$ the radial mode number, of which the Q-factor is $1230$, in a concentric circular layered cavity with refractive indices $n_i = 1.8$ and $n_o = 1.0$, respectively, when $r_i = 0.6$ and $\epsilon = 0$. This concentric circular layered cavity is the same as a circular cavity with a radius of $r_i$ and a refractive index of $n_i$. As we increase $\epsilon$, the WGM in the concentric circular layered cavity changes into a cWGM in a lima\c{c}on-shaped DLTC.

\begin{figure}[tb]
\centering
\includegraphics[width=\linewidth]{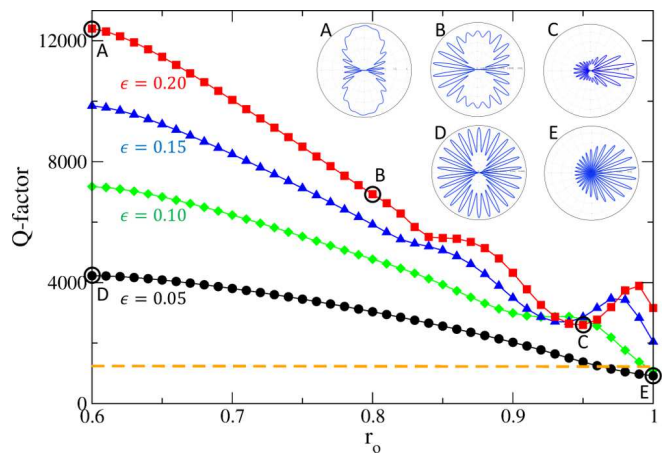}
\caption{Q-factors of the resonant modes in double-layered lima\c{c}on-shaped transformation cavities as a function of $r_o$ when $\epsilon=0.05$ (black circles), $\epsilon=0.1$ (green diamonds), $\epsilon=0.15$ (blue triangles), and $\epsilon=0.2$ (red squares). The insets display selected far-field intensity patterns. The orange dashed line represents the Q-factor of the whispering gallery mode with $(m,l)=(13,1)$ in a circular cavity with $n_0=1.8$.}
\label{fig3}
\end{figure}

A physical space consists of a gradient index cavity and the air, while a WT space should consist of a gradient index cavity and a gradient index of an infinite whole space which cannot be realized. As the size of the outer layer increases, resonant modes in the TC become closer to those in a WT space; for example, a cWGM in a TC approaches the corresponding WGM in a gradient index cavity in a WT space, which is equivalent to a WGM in a homogeneous circular cavity in an OV space. It should be noted that this crossover between a gradient index cavity in a WT space and a TC is not limited to the cWGM, i.e., it is not the mode properties but rather the system characteristics.

{\it Q-factors as a function of $r_o$ -}
To show the crossover from a TC to a gradient index cavity in a WT space, we fix the deformation of the cavity $\epsilon$ and the size of the inner layer $r_i$ but change the outer layer boundary.
In the case of a TC without an outer layer, $r_{o} = 0.6$, as $\epsilon$ increases, the Q-factors of the corresponding modes increase in the region of $\epsilon < 0.2$ because of the effect of increasing the refractive index inside the cavity due to the distortion of the length scale by the conformal mapping \cite{Ryu21}. As we increase the size of outer layer $r_o$, the Q-factors of the resonant modes monotonically decrease if $r_o \lesssim 0.8$, as shown in Fig.~\ref{fig3}, because the outer layer effectively reduces the effect of higher refractive index ratios at the inner layer boundary originating from the distortion of the length scale inside the cavity by conformal mapping. If $r_o \gtrsim 0.8$, the Q-factors show fluctuations as $r_o$ increases since a sufficiently large outer layer plays the role of an additional resonator. This kind of transition between monotonic and oscillatory behaviors depending on the size of an additional resonator where the resonances are coupled to open channels is well known in coupled microcavities \cite{Ryu06}. 
If we can infinitely increase $r_o$ like in the case of a WT space, the Q-factor approaches that of the corresponding WGM in a circular cavity, such as the dashed line in Fig.~\ref{fig3}, because the two resonant modes in OV and WT spaces are equivalent. It is noted that the Q-factors of resonant modes in slightly deformed cases such as $\epsilon=0.05$ can be smaller than that of the corresponding WGM in a circular cavity in an oscillatory region.

\begin{figure}[tb]
\centering
\includegraphics[width=\linewidth]{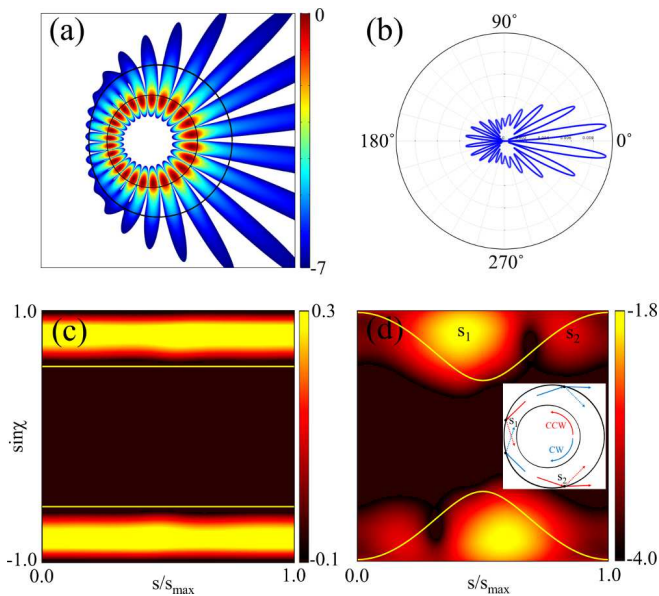}
\caption{(a) Near-field and (b) far-field intensity patterns of a resonant mode in a double-layered transformation cavity when $r_i = 0.6$, $r_o = 0.95$, and $\epsilon = 0.2$. The corresponding Husimi functions are shown for (c) the inner and (d) the outer layer boundary, where $s$ is the boundary coordinate in the reciprocal virtual space \cite{Kim18}. The yellow lines represent the critical lines for total internal reflection, i.e., the ratios of the refractive indices at the dielectric interfaces of the inner layer boundaries. In the inset, straight (dotted) arrows inside the cavity denote the boundary positions and incident (reflection) angles corresponding to $s_{1}$ and $s_{2}$. The arrows outside cavity denote the boundary positions and refractive angles of the dominant emissions. The red and blue arrows represent the counter-clockwise ($\sin{\chi} > 0$) and clockwise ($\sin{\chi} < 0$) components, respectively.}
\label{fig4}
\end{figure}

{\it Far-field intensity patterns -}
For the different $r_o$, the near-field patterns inside the inner layer are not distinct from each other [Fig.~\ref{fig4} (a)], i.e., they are cWGMs in the inner layers. On the other hand, the far-field patterns show clearly different emission directionalities, as shown in Fig.~\ref{fig3}. In the case of a single transformation cavity with sufficiently large $\epsilon$, the far-field patterns of cWGMs show bidirectional emissions, which are explained by the universal tunneling mechanism related to the boundary shapes and the inhomogeneous refractive index profiles determined by the given conformal mapping \cite{Kim16}. As $r_o$ increases, the emission directionalities of the far-field patterns change from bidirectional to unidirectional via mixed-direction for large $\epsilon$ cases. Additionally, the nearly isotropic emissions of the far-field patterns change into unidirectional emissions for small $\epsilon$ cases such as $\epsilon=0.05$.

{\it Near-field intensity patterns and Husimi functions -}
The remarkable changes of the far-field intensity patterns as $r_o$ increases are caused by the modification of the outer layer boundary, not by the significant changes of the near-field intensity patterns around the perimeter of the inner layer. Figure~\ref{fig4} (c) and (d) show Husimi functions at the inner and the outer layer boundary, respectively, of cWGMs forming along the inner layer boundary. The Husimi function is a phase space representation of intracavity wave intensity and is defined at the dielectric interfaces by the overlap of the boundary wave function of a resonant mode with a Gaussian wave packet on the cavity boundary in a phase space \cite{Hen03,Lee05,Kim18}.
In a conventional single TC, the cWGM has varying tunneling emission according to cavity boundary position since the ratio of the refractive indices at the dielectric interfaces of the cavity boundaries is given as a function of boundary position. Tunneling emission mainly occurs at the position where the ratio is minimum, and light is emitted along the tangent direction at this position. In the DLTC, however, the cWGM exhibits a nearly isotropic tunneling emission at the inner layer boundary since the ratio of the refractive indices at the dielectric interface between the inner and outer layers is constant, $n_0$, independent of boundary position as shown in Fig.~\ref{fig4} (c). In Fig.~\ref{fig4} (a), the modes are better confined in the left part of the outer layer because of the large contrast between the refractive indices of the outer layer and air. A localized high intensity of the Husimi function of the resonant mode above the critical line for total internal reflection around $s_1$ of Fig.~\ref{fig4} (d) confirms well-confined light. In contrast, the modes are emitted through the right part of the outer layer boundary because the ratio of the refractive index between the outer layer and air at that region is small. The second highest intensity of the Husimi function around $s_2$ mainly locates below the critical line for total internal reflection, and this produces the refractive emission in the right part of the outer layer boundary shown in the inset of Fig.~\ref{fig4} (d). This mechanism achieves unidirectional emission in the DLTC.

\begin{figure}[t]
\centering
\includegraphics[width=\linewidth]{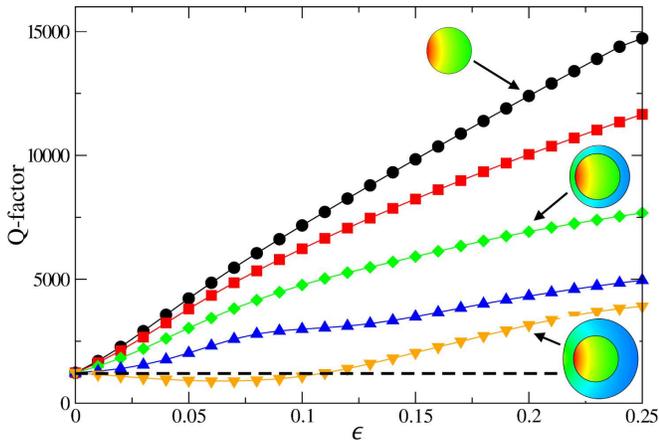}
\caption{Q-factors of the resonant modes in double-layered lima\c{c}on-shaped transformation cavities as a function of $\epsilon$ for $r_o = 0.6$ (black circles, transformation cavity without an outer cavity), $0.7$ (red squares), $0.8$ (green diamonds), $0.9$ (blue triangles), and $1.0$ (orange inverted triangles) when $r_i=0.6$. The black dashed line represents the Q-factor of the whispering gallery mode with $(m,l)=(13,1)$ in a circular cavity with $n_0=1.8$. The insets depict the boundary shapes and refractive index profiles of single and double-layered limaçon-shaped transformation cavities. The color scales of the refractive index profiles are the same as in Fig.~\ref{fig1}.}
\label{fig5}
\end{figure}

{\it Q-factors as a function of $\epsilon$ -} Finally, we study the Q-factors of the modes as a function of $\epsilon$ for different $r_o$ when $r_i = 0.6$ (Fig.~\ref{fig5}). In the case of a TC without an outer cavity ($r_o = 0.6$), as $\epsilon$ increases, the Q-factors of the corresponding modes increase as discussed above. In some cases, if $\epsilon$ increases further, the Q-factors show a decreasing trend \cite{Kim16}. If we can infinitely increase $r_o$, like in the case of conventional TO, the Q-factor does not depend on the change of $\epsilon$ such as the dashed line in Fig.~\ref{fig5} because the two resonant modes in OV and WT spaces are equivalent. If we increase $r_o$, the rate of Q-factor increase decreases, and the Q-factors are mostly maintained regardless of $\epsilon$ when $r_o = 1.0$.

\section{Summary}

As an alternative model of a transformation cavity having a gradually varying refractive index profile in the interior region of the cavity, a double-layered transformation cavity consisting of inner and outer layers with gradually varying refractive index profiles has been proposed. We have studied the Q-factors, near- and far-field intensity patterns, and Husimi functions of the resonant modes in double-layered transformation cavities. As the outer layer becomes larger, the Q-factors and near-field intensity patterns of the resonant modes in the newly proposed cavities become closer to those in a cavity in the wholly transformed space of transformation optics. As the outer layer boundary is changed without modifications to the inner layer, the far-field patterns of the modes in the cavities can be controlled, while the near-field patterns in the inner layers do not change. We expect that such a control scheme of the emission directionality of the resonant modes without significant deformation of the mode pattern only by modification of the outer layer without changing the inner layer paves the way to design a versatile cavity through the suggested double-layered transformation cavity. As a similar example, hybrid optical microcavities comprising an inner transformation cavity and an outer homogeneous deformed cavity have recently been proposed to extract unidirectional narrow-beam emission from high-Q whispering gallery modes \cite{Lim21}.

\section*{acknowledgments}
The author thanks J. Cho, I. Kim, S. Rim, and M. Choi for helpful discussions. We acknowledge financial support from the Institute for Basic Science in the Republic of Korea through the project IBS-R024-D1.

\end{document}